\documentclass[preprint, showkeys]{revtex4-1}
\pdfoutput=1
\usepackage[utf8]{inputenc}
\usepackage[english]{babel}
\usepackage{graphicx}

\begin{document}
\newcommand{\Mo}[1]{$^{#1}\mathrm{Mo}$}    
    
\title{Simulation of the liquid targets for molybdenum-99 production}
\author{D.V. Fedorchenko}
\email[Corresponding author: ]{fdima@kipt.kharkov.ua}

\author{M.A. Khazhmuradov}

\author{Y.V. Rudychev}
\affiliation{National Science Center ``Kharkov Institute of Physics and
	Technology'', Kharkiv, Ukraine}    

\begin{abstract}
The efficiency of \Mo{99} nuclei trapping by clinoptilolite particles using Monte Carlo simulation was studied. The simulation showed the carrier particle traps almost all of the incident \Mo{99} nuclei for the photon energies 12-18 MeV. The ratio of the \Mo{99} nuclei reaching the carrier to the total number of \Mo{99} nuclei escaping the nanoparticle is below 1.5\% in for the photon energies 12-18~MeV. High specific activity of produced \Mo{99} nuclide requires optimization of the clinoptilolite carrier particles concentration in the suspension. 
\end{abstract}



\maketitle

\section{Introduction}

The isotope \Mo{99} plays prominent role in the nuclear medicine. $\beta -$ decay of \Mo{99} generates ${}^{99\text{m}}$Tc nuclide which is the primary radionuclide used in single photon emission computed tomography (SPECT). The growing market demands inspire the permanent interest to the development of safe and efficient methods for \Mo{99} nuclide production. 

Modern commercial production of \Mo{99} nuclide uses nuclear research reactors, where \Mo{99} is generated as a fission product of ${}^{235}$U isotope. Recently, the issues related to non-proliferation and environmental protection gave rise to the accelerator based methods of \Mo{99} nuclide production \cite{Dikiy2000,Starovoitova2014,Mangera2015,Fujiwara2017}. However, despite the remarkable progress of this technology two main problems remain unsolved: low specific activity of the produced nuclide and high heat loads of the irradiated target.

Low specific activity of \Mo{99} produced using accelerators is a consequence of the low cross section of the corresponding nuclear reaction. Table~\ref{tab:cross-sections} shows that cross sections of nuclear reactions used for \Mo{99} accelerator-based production are two orders of magnitude lower than effective cross section of ${}^{235}$U fission. Thus, higher specific activity could be achieved either by extended irradiation time or by higher fluence through the irradiated target. However, \Mo{99} half-life of approximately 66 hours limits the irradiation time and high heat loads of the production target impose constraints on the incident particles fluence. 

Another problem connected with accelerator based production of \Mo{99} is extraction process that requires complete dissolution of the irradiated target. Low specific activity of the produced molybdenum leads to the high production expenses, as target recycling is impossible and a considerable amount of wastes is produced.

High specific activity of the produced molybdenum could be achieved using the kinematic recoil method \cite{Starovoitova2014,Dikiy-nano}. This method uses trapping of the recoil molybdenum nuclei escaping the target by the small-sized carrier particles. The possible implementation of this approach is a liquid target containing suspension of molybdenum nanoparticles and clinoptilolite carrier particles [10]. In this case recoil molybdenum nuclei are trapped by the 100-200 nm sized carrier particles. After the irradiation carrier particles containing \Mo{99} are filtered out from the suspension and target can be reused with the new portion of carrier particles. 

\begin{table}
    \caption{\label{tab:cross-sections}Cross sections of the \Mo{99} producing reactions.}
    \begin{ruledtabular}
        \begin{tabular}{llll}
Reaction & Projectile energy	& Cross section, barn &	Ref.\\
\multicolumn{4}{c}{Reactor production}\\
$^{235}$U(n,f)\Mo{99}   &	0.025~eV &	582 (\Mo{99} yield $\approx$6\%)& \cite{SHIBATA2011,Koning2006}\\
\Mo{98}(n,$ \gamma $)\Mo{99}  & 0.025~eV &	0.13                            & \cite{Chadwick2011}  \\
\multicolumn{4}{c}{Accelerator production}\\
\Mo{100}($ \gamma $,n) \Mo{99}&	16 MeV	 &  0.16                            & \cite{Beil1974} \\
\Mo{100}(p,pn)\Mo{99}         & 35 MeV   &  0.16                            & \cite{CERVENAK201632} \\
\Mo{100}(p,2n) ${}^{99\text{m}}$Tc & 14 MeV& 0.24                           &\cite{CERVENAK201632}\\
        \end{tabular}
    \end{ruledtabular}
\end{table}

In the previous paper \cite{Dikiy2016} \Mo{99} production method using \Mo{100}($ \gamma $,n)\Mo{99} photonuclear reaction and simulated process of recoil nuclei transport in the molybdenum nanoparticle was considered. The subject of the present study was the process of recoil molybdenum nuclei trapping by the carrier particle. The process of recoil nuclei transport inside the mother nanoparticle, the ambient liquid and carrier particle was simulated in order to estimate the trapping efficiency for various energies of the incident photons. Transport simulations were performed using Geant4 simulation toolkit~\cite{Geant1,Geant2}, while for the reasons of computational efficiency the energy spectra of photonuclear reaction were calculated using nuclear reaction code TALYS \cite{Talys1}.

\section{Methods}
\subsection{Simulation of the photonuclear reaction}

\begin{figure}
    \centering
    \includegraphics{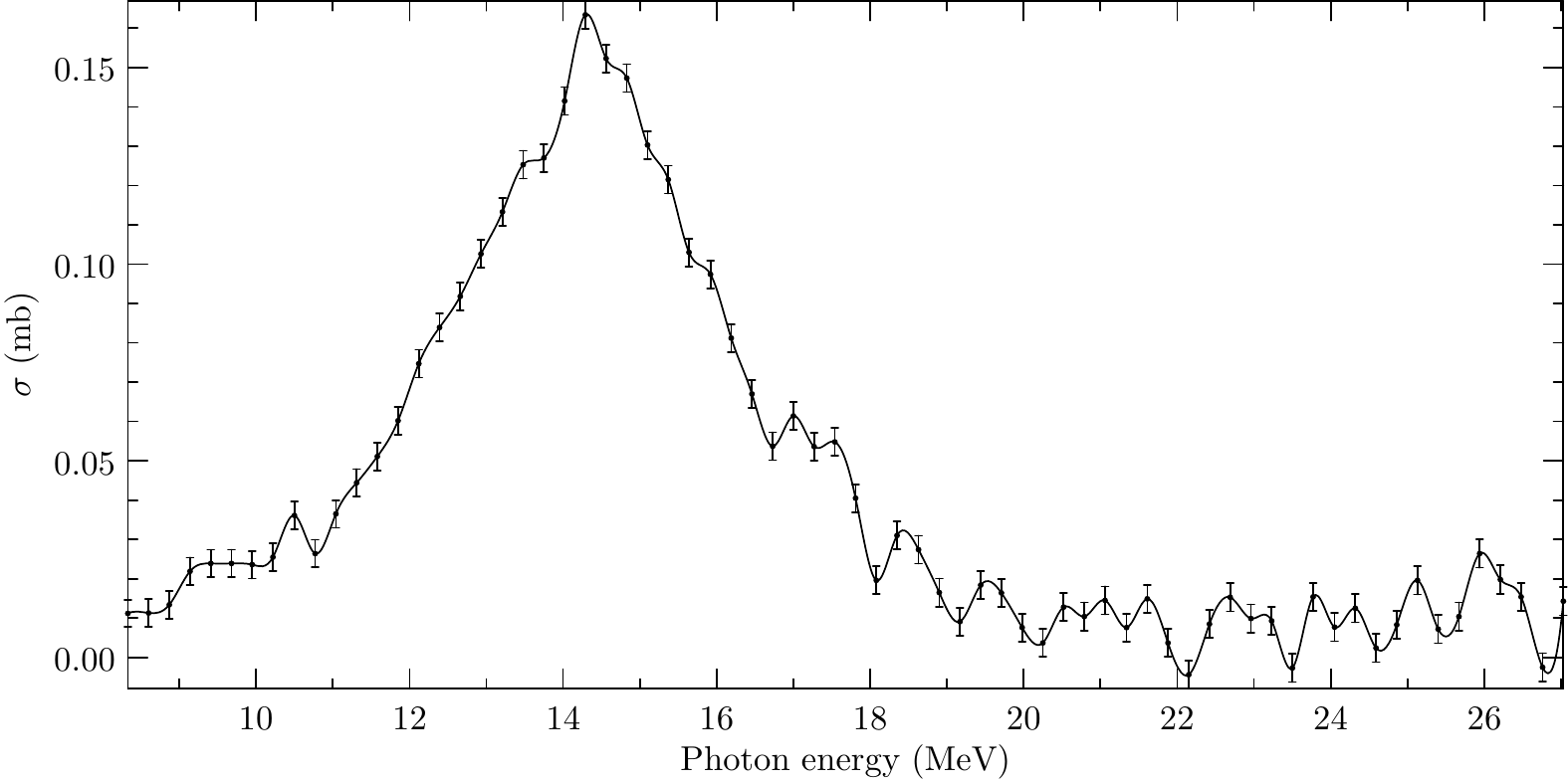}
    \caption{\label{fig:ris1}Cross section of the \Mo{100}($\gamma$,n)\Mo{99} reaction (data taken from \cite{CERVENAK201632}).}    
\end{figure}

The energy spectrum of the \Mo{100}($ \gamma $,n)\Mo{99} reaction defines 
initial kinetic energy of the recoil \Mo{99} nucleus, and consequently, its 
mean free path in mother nanoparticle, surrounding liquid and carrier particle. 
The cross section of this reaction has the broad peak around 14 MeV due to the 
Giant Dipole Resonance (GDR) effect (Figure~\ref{fig:ris1}). The GDR peak 
defines the energy interval of incident photons for efficient production of 
\Mo{99} from 12 to 18 MeV. For such photon energies energy spectra of outgoing 
neutrons and \Mo{99} recoil nuclei are described with sufficient accuracy by 
the evaporation model~\cite{Weisskopf1937}. Geant4 toolkit implementation of 
this model for the Monte Carlo calculations uses the approach described in the 
work \cite{Dostrovsky1959}.

Small cross section of photon-nucleus interaction makes direct simulation of 
the photonuclear reaction in the single molybdenum nanoparticle rather 
inefficient. This problem has several possible solutions. For example, 
non-analog Monte Carlo simulation uses artificially increased photonuclear 
reaction cross section or interaction probability \cite{Koning2006}. In this 
paper another approach was used: the spectrum of the photonuclear reaction was 
calculated beforehand using nuclear reaction code TALYS 1.8 \cite{Talys1} and 
then used for Monte Carlo simulations of the recoil nuclei transport. TALYS is 
software package for simulation of the nuclear reaction in the wide energy 
range. It contains reliable models for the various types of nuclear reactions 
including photonuclear.

\begin{figure}
    \centering
    \includegraphics{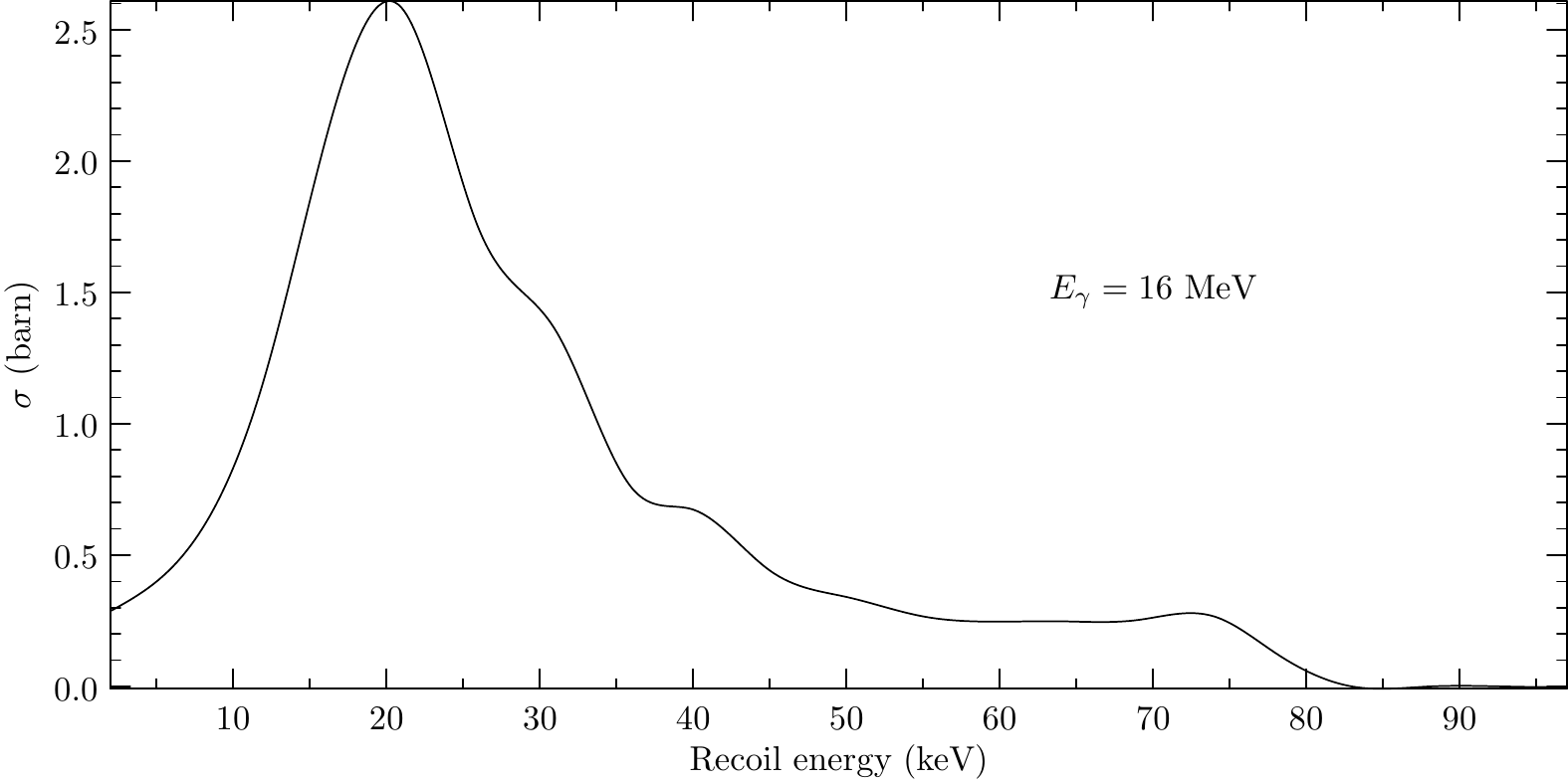}
    \caption{\label{fig:ris2}Calculated spectrum of \Mo{99} recoil nuclei for 16~MeV incident photons (TALYS~1.8).}    
\end{figure}

Figure~\ref{fig:ris2} shows the calculated \Mo{99} recoil nuclei spectrum for 16~MeV incident photons. The recoil spectrum has pronounced peak around 20~keV and rather long high-energy tail up to recoil nuclei energies of about 80~keV. 
Position of the peak depends on the excitation energy of the compound nucleus 
created after the photon capturing, and shifts to the higher energies as the 
energy of the incident photon increases. The calculated spectrum shows that the 
major fraction of \Mo{99} recoil nuclei has the energies from 10 to 40~keV. 
This energy region defines the effective range of the recoil nucleus in the 
ambient liquid and in the carrier particles.

\subsection{Simulation of the recoil nuclei transport}

For the simulation of \Mo{99} recoil nuclei transport Geant4 toolkit~\cite{Geant1,Geant2} version 10.2.03  was used. This freely distributed simulation toolkit is developed and maintained by the Geant4 Collaboration. Geant4 is a software framework for Monte Carlo simulations for high energy physics applications. It is comprised of numerous C++ classes describing various aspects of particle transport: particles descriptions, media properties, models of physical processes, problem geometry, and etc. The open modular architecture of the framework simplifies addition of the new models of physical processes. 

The simulation setup for \Mo{99} nuclei trapping included 40~nm sized nanoparticle and 160~nm carrier particle immersed in the ethylene glycol at the distance of several tens of nanometers from each other. As nanoparticle materials pure \Mo{100} and molybdenum oxide (MoO$_3$) were considered. \Mo{100} is the preferred material because of a relatively small number of accompanying molybdenum nuclei produced~\cite{Dikiy2016}, however it is rather expensive. Molybdenum oxide MoO$_3$ provides the cheaper alternative. It has the lower density which favors escaping of the recoil \Mo{99} nuclei, but at the same time large amount of accompanying molybdenum isotopes (\Mo{92}, \Mo{93}, \Mo{94}, \Mo{95}, \Mo{96}, \Mo{97}, \Mo{98}, \Mo{100}) are produced. The carrier particle material was clinoptilolite, which is a natural zeolite with chemical formula (Na,K,Ca)$_{2-3}$A$_{l3}$(Al,Si)$_2$Si$_{13}$O$_{36}\cdot 12$H$_2$O.

Despite the simple geometry of the model it imposes essential restrictions on the applicable models of physical processes. Namely, as the range of the recoil \Mo{99} nucleus in molybdenum is about 5-10~nm \cite{Starovoitova2014} only a small number collisions occur along the recoil nuclei path. Thus high precision model is required for simulation of recoil nucleus elastic scattering to obtain the physically consistent results.

For actual simulation accurate model for elastic ion scattering developed by Mendenhall and Weller \cite{Mendenhall2005420} was used. This model implements the classical description of the ions elastic scattering on the screened Coulomb potential. Benchmark calculations presented in \cite{Mendenhall2005420} showed the reasonable agreement of this model with experimental data for $ \alpha $-particles scattering on 100~nm-thick foils, and for simulation of boron implantation.

Geant4 package contains G4ScreenedNuclearRecoil class that implements single scattering algorithm \cite{Mendenhall2005420}. For the simulation the constructor of the corresponding class was added to the standard Geant4 physical list used for the simulation of electromagnetic processes. This list contains models for Compton scattering, photoelectric effect and pair production. For the reasons of computational efficiency hadronic interactions were completely omitted because of their low probability for the recoil nuclei with energies of several tens of kilovolts.

\section{Results and discussions}

The first quantity calculated was the average range of the escaped \Mo{99} nuclei in ethylene glycol. This value defines the characteristic distance between nanoparticle and carrier for recoil nuclei trapping. However, average range corresponds to the center of the recoil nuclei path lengths distribution. If one considers the number of nuclei capable to reach the carrier, the median of this distribution is more suitable parameter as it establishes the distance that more than a half of escaped particles surpass.

For the calculations of range and median the simple model consisted of the spherical 40~nm nanoparticle immersed in the ethylene glycol was used. The spectra of the recoil nuclei were calculated using TALYS~1.8 code and then were used in subsequent Monte Carlo simulation using Geant4. Figure 3 shows the average range and median of the \Mo{99} recoil nuclei path lengths distribution. It follows that for the photons with energies in GDR region (12-18~MeV) the median of the path length distribution is about 30-40 nanometers. This value defines the distance between nanoparticle and carrier for efficient trapping of the recoil nuclei. The smaller distances favor recoils trapping; however distances between molybdenum particles and carrier particles below 10 nm are achievable only for very concentrated suspension. In this case a large number of the escaping nuclei are trapped by the neighbor molybdenum nanoparticles and the number of \Mo{99} nuclei trapped by the carrier particles decreases.

\begin{figure}
    \centering
    \includegraphics{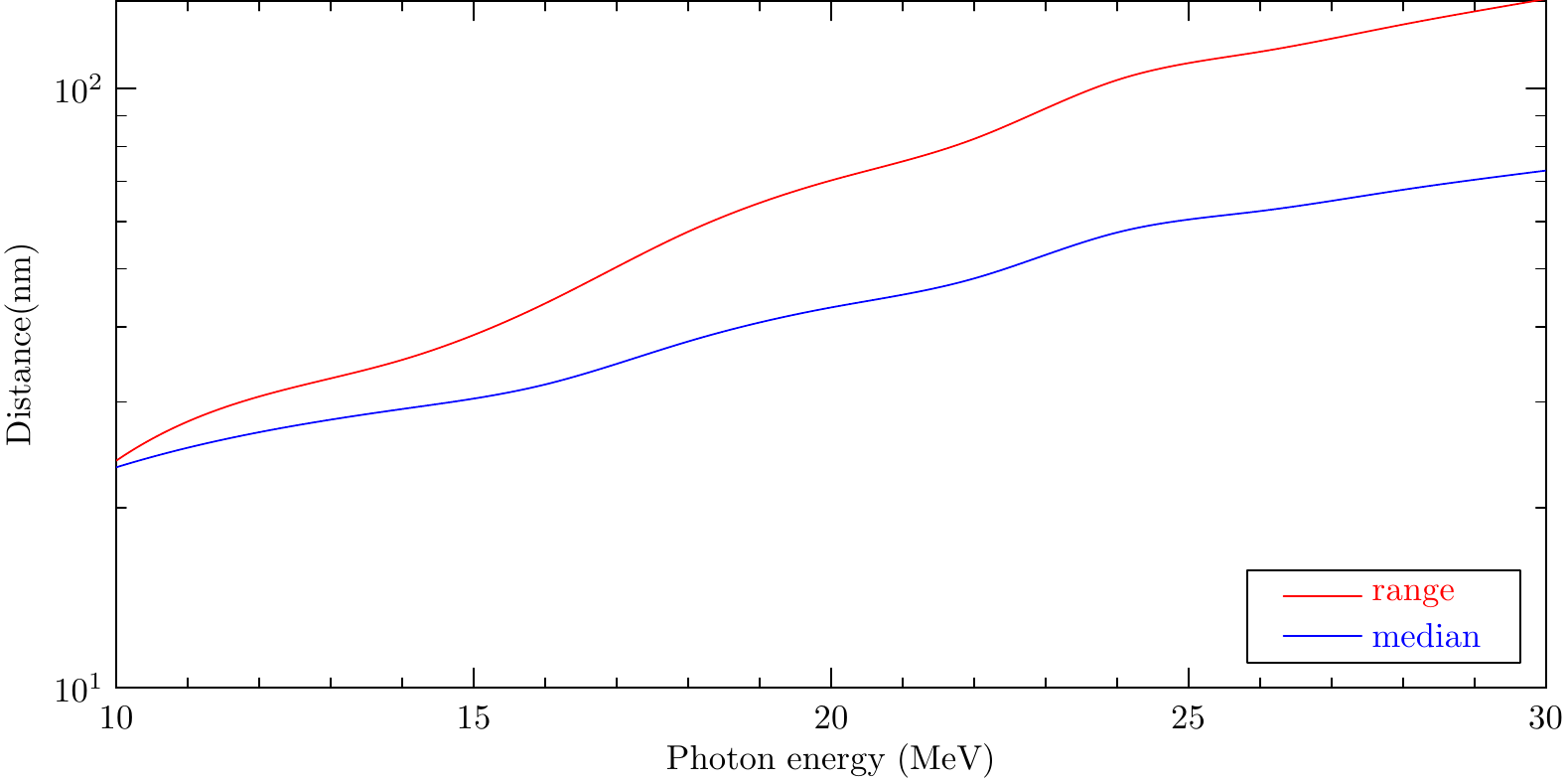}
    \caption{\label{fig:ris3}\Mo{99} recoil nuclei path length distribution: average range and median.}    
\end{figure}

For the higher energies median of the path length distribution exceeds the 100~nm value. However, high-energy photons constitute only a small fraction of bremsstrahlung spectrum used for photonuclear production of \Mo{99} \cite{Dikiy2016} and cross section of photonuclear reaction at such energies is small (see Figure~\ref{fig:ris1}). Thus, only a small number of high-energy recoil nuclei is actually produced.

\begin{figure}
    \centering
    \includegraphics[width=\linewidth]{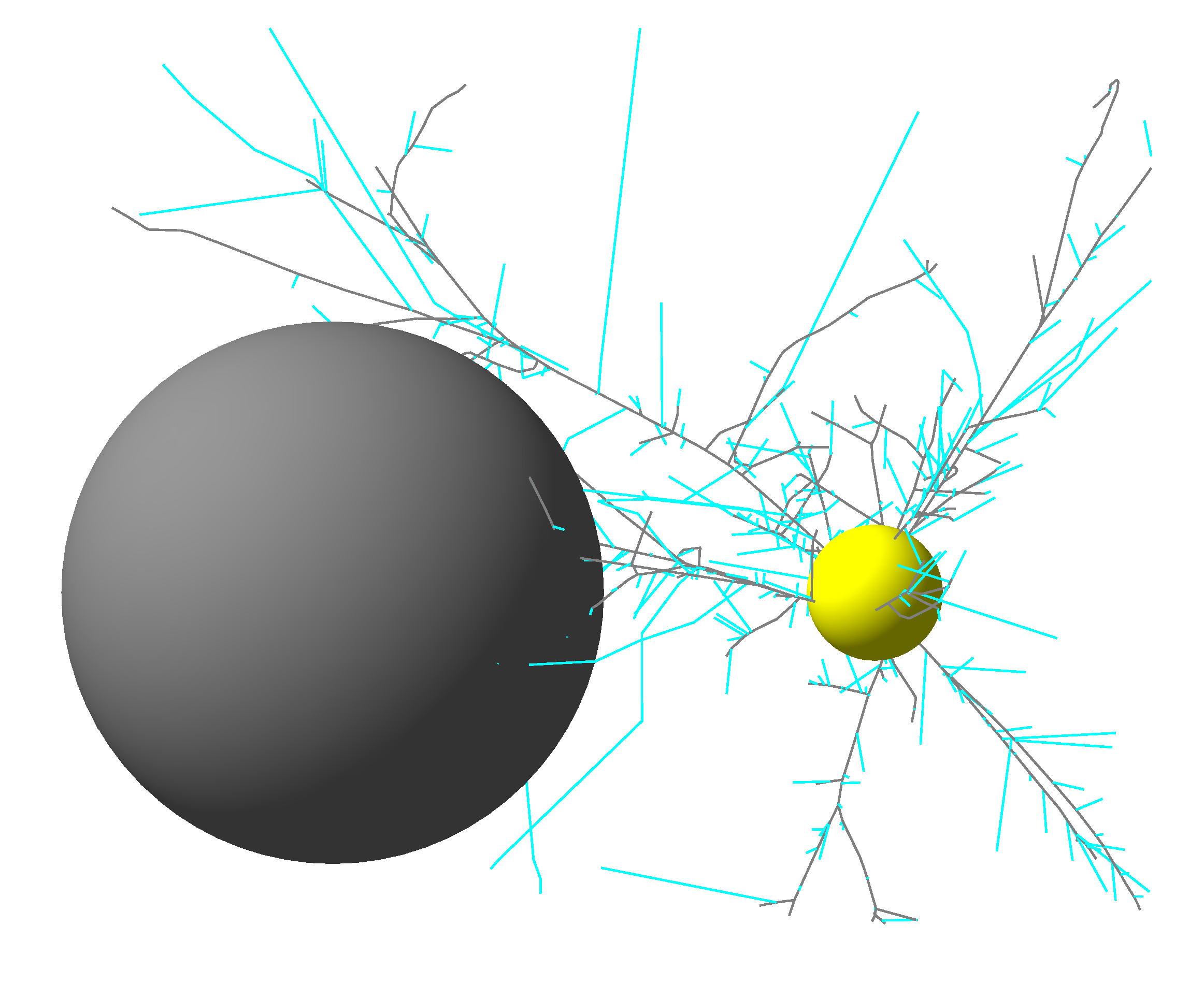}
    \caption{\label{fig:ris4}Geant4 model used for simulation of \Mo{99} trapping in the carrier particles.}    
\end{figure}

Calculations of the average range and median were used to define the appropriate interval of the interparticle distances for simulation of recoil nuclei trapping. Hereafter term interparticle distance means the distance between the surfaces of molybdenum nanoparticle and carrier particle. For the simulation three values of interparticle distances were chosen: 20, 40 and 60~nm. Figure~\ref{fig:ris4} shows the corresponding Geant4 model containing 40 nm nanoparticle (yellow) and 160 nm carrier particle (gray).

\begin{figure}
    \centering
    \includegraphics{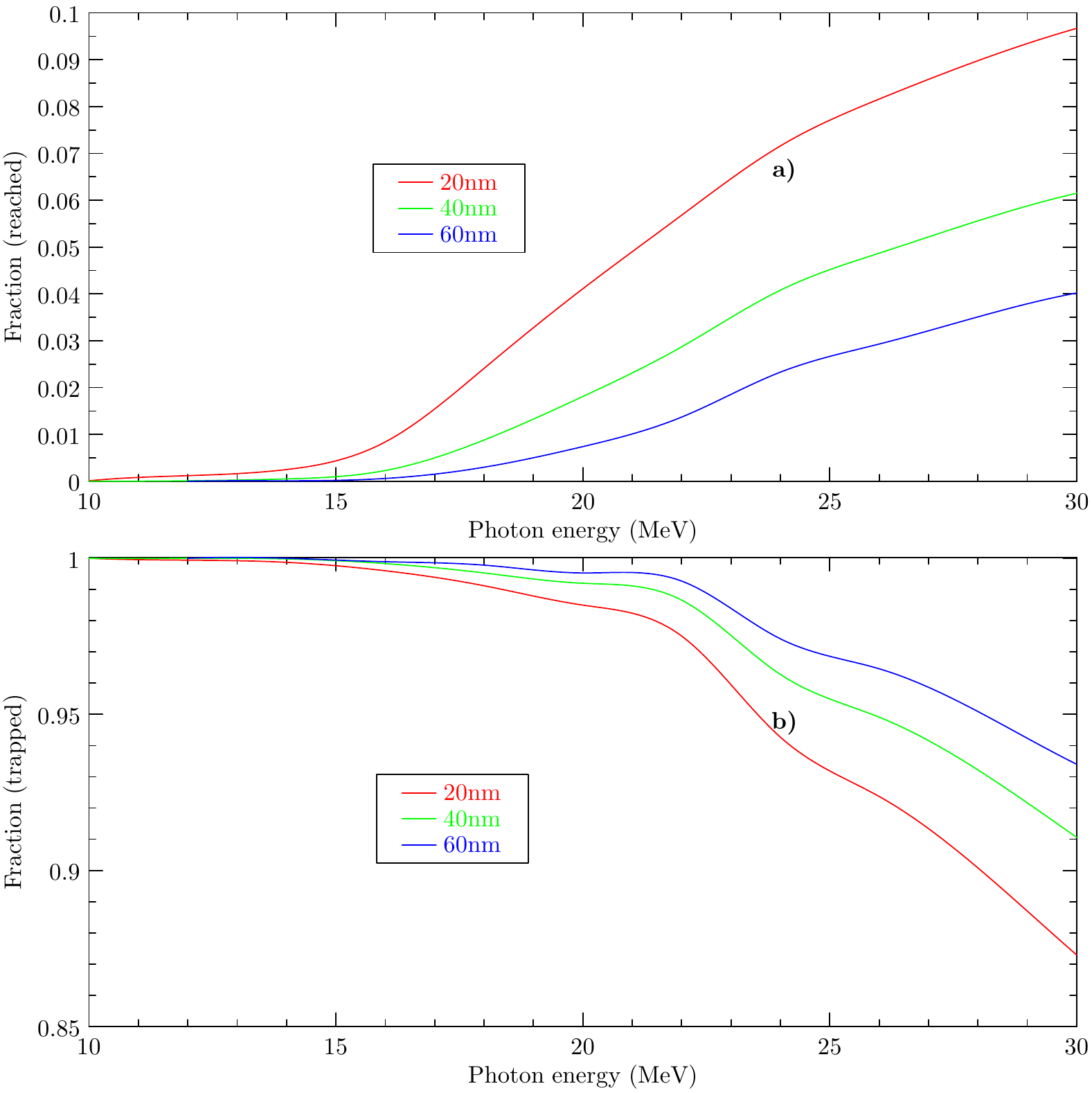}
    \caption{\label{fig:ris5}Fraction of the \Mo{99} recoil nuclei reached the carried (a) and trapped (b) for \Mo{100}.}    
\end{figure}

\begin{figure}
    \centering
    \includegraphics{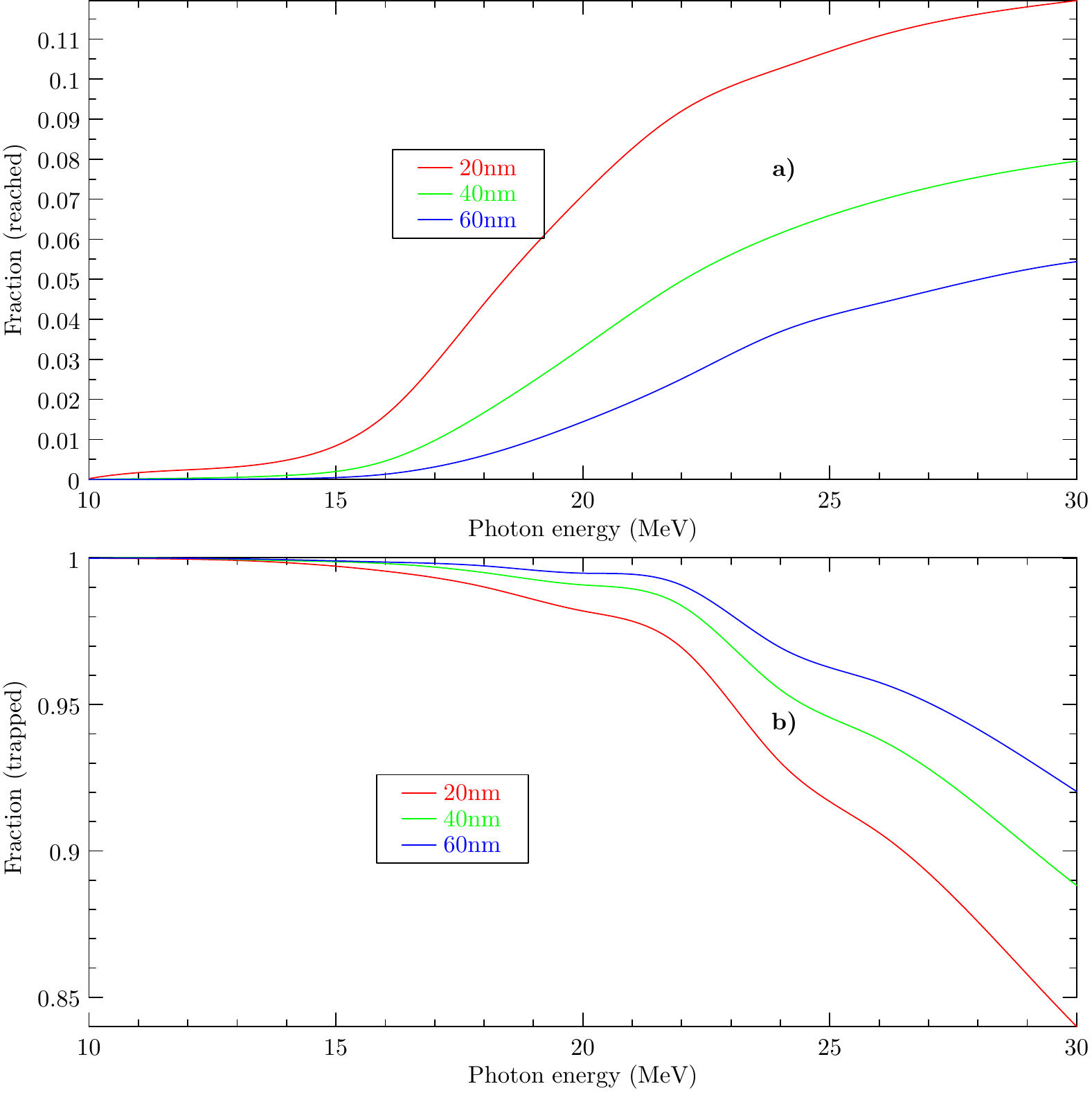}
    \caption{\label{fig:ris6}Fraction of the \Mo{99} recoil nuclei reached the carried (a) and trapped (b) for MoO$_3$.}    
\end{figure}

From Figures~\ref{fig:ris5}a and \ref{fig:ris6}a it follows that fraction of the recoil \Mo{99} nuclei that reaches \Mo{100} (Figure~\ref{fig:ris5}a) and MoO$_3$ (Figure~\ref{fig:ris5}b) carrier particles for incident photons with energies in GDR region is about 0.5-1.5\%, and only for high energy photons exceeds the value of 10\%. Figures~\ref{fig:ris5}b and \ref{fig:ris6}b show that almost all these nuclei are trapped by the carrier particle. Thus, the resulting fraction of the initial recoil nuclei trapped by the carrier is below 1.5\%. The slightly lower trapping ratio for MoO$_3$ nanoparticles is attributed to the lower density of the molybdenum oxide (4.69 g/cm$^3$) compared to pure \Mo{100} (10.32 g/cm$^3$).

From our previous calculations (see \cite{Dikiy2016} and Figure~8 herein) it follows that for photons with energies in the GDR region the escape ratio of the \Mo{99} recoils is 10-40\% for \Mo{100} and 18-60\% for MoO$_3$. Considerably low fraction of the trapped recoil nuclei obtained in this work results from two main factors: noticeable stopping power of the recoil nuclei in ethylene glycol and model geometry. 

The influence of the first factor could be easily understood using the Figure 3 showing that for 16~MeV photons the median of path length distribution is about 40~nm. Thus, the number of recoil nuclei capable to reach the carrier particle is halved as a result of slowing down in the nanoparticle material and liquid media. 

The geometry factor limits constrains the possible velocity directions recoil nuclei escaping from nanoparticle to those close to the direction from nanoparticle to carrier particle. The elementary geometrical considerations give the value of the corresponding solid angle of 0.83, or 6.6\% of the full 4$ \pi $ solid angle for 40~nm interparticle distance. Combining this with 30-40\% loss of the initial recoil nuclei in the nanoparticle and 50\% loss along the path to the carrier particle one comes to the crude estimation of fraction of \Mo{99} that reaches carrier of 1.0-1.5\%. This value is close to the results of Monte Carlo simulation (see Figures~\ref{fig:ris5} and \ref{fig:ris6}).

The above considerations show that amount of the collected \Mo{99} nuclei could be increased if each carrier particle neighbors the several molybdenum nanoparticles placed at 20-40 nm distance. In fact, obtaining the carrier particles with the high specific activity of \Mo{99} requires optimal concentrations of molybdenum nanoparticles and carrier particles. However, formal solution of this optimization problem is highly complicated and further experimental studies are required to evaluate the optimum concentrations. 

\section{Conclusions}
Monte Carlo simulations of the \Mo{99} radioisotope production process using molybdenum nanoparticles and clinoptilolite carrier particles for trapping the recoil molybdenum nuclei were performed. The calculations showed that single carrier particle collects only 1-1.5\% of the recoil \Mo{99} nuclei. The low value of the trapping ratio appears to be due to geometrical reasons and could be increased by using higher concentration of the carrier particles. However, this requires additional studies to find out the optimal concentration of the carriers for efficient trapping of the produced \Mo{99} nuclei.

\bibliography{science}

\begin{thebibliography}{17}%
\makeatletter
\providecommand \@ifxundefined [1]{%
 \@ifx{#1\undefined}
}%
\providecommand \@ifnum [1]{%
 \ifnum #1\expandafter \@firstoftwo
 \else \expandafter \@secondoftwo
 \fi
}%
\providecommand \@ifx [1]{%
 \ifx #1\expandafter \@firstoftwo
 \else \expandafter \@secondoftwo
 \fi
}%
\providecommand \natexlab [1]{#1}%
\providecommand \enquote  [1]{``#1''}%
\providecommand \bibnamefont  [1]{#1}%
\providecommand \bibfnamefont [1]{#1}%
\providecommand \citenamefont [1]{#1}%
\providecommand \href@noop [0]{\@secondoftwo}%
\providecommand \href [0]{\begingroup \@sanitize@url \@href}%
\providecommand \@href[1]{\@@startlink{#1}\@@href}%
\providecommand \@@href[1]{\endgroup#1\@@endlink}%
\providecommand \@sanitize@url [0]{\catcode `\\12\catcode `\$12\catcode
  `\&12\catcode `\#12\catcode `\^12\catcode `\_12\catcode `\%12\relax}%
\providecommand \@@startlink[1]{}%
\providecommand \@@endlink[0]{}%
\providecommand \url  [0]{\begingroup\@sanitize@url \@url }%
\providecommand \@url [1]{\endgroup\@href {#1}{\urlprefix }}%
\providecommand \urlprefix  [0]{URL }%
\providecommand \Eprint [0]{\href }%
\providecommand \doibase [0]{http://dx.doi.org/}%
\providecommand \selectlanguage [0]{\@gobble}%
\providecommand \bibinfo  [0]{\@secondoftwo}%
\providecommand \bibfield  [0]{\@secondoftwo}%
\providecommand \translation [1]{[#1]}%
\providecommand \BibitemOpen [0]{}%
\providecommand \bibitemStop [0]{}%
\providecommand \bibitemNoStop [0]{.\EOS\space}%
\providecommand \EOS [0]{\spacefactor3000\relax}%
\providecommand \BibitemShut  [1]{\csname bibitem#1\endcsname}%
\let\auto@bib@innerbib\@empty
\bibitem [{\citenamefont {Dikiy}\ \emph {et~al.}(2000)\citenamefont {Dikiy},
  \citenamefont {Dovbnya}, \citenamefont {Lyashko}, \citenamefont {Medvedeva},
  \citenamefont {Tur}, \citenamefont {Uvarov}, \citenamefont {Fedorets},
  \citenamefont {Pashchuk},\ and\ \citenamefont {Evseev}}]{Dikiy2000}%
  \BibitemOpen
  \bibfield  {author} {\bibinfo {author} {\bibfnamefont {N.}~\bibnamefont
  {Dikiy}}, \bibinfo {author} {\bibfnamefont {A.}~\bibnamefont {Dovbnya}},
  \bibinfo {author} {\bibfnamefont {Y.}~\bibnamefont {Lyashko}}, \bibinfo
  {author} {\bibfnamefont {E.}~\bibnamefont {Medvedeva}}, \bibinfo {author}
  {\bibfnamefont {Y.}~\bibnamefont {Tur}}, \bibinfo {author} {\bibfnamefont
  {V.}~\bibnamefont {Uvarov}}, \bibinfo {author} {\bibfnamefont
  {I.}~\bibnamefont {Fedorets}}, \bibinfo {author} {\bibfnamefont
  {S.}~\bibnamefont {Pashchuk}}, \ and\ \bibinfo {author} {\bibfnamefont
  {I.}~\bibnamefont {Evseev}},\ }\href@noop {} {\bibfield  {journal} {\bibinfo
  {journal} {Probl. At. Sci. Technol.}\ }\textbf {\bibinfo {volume} {36}},\
  \bibinfo {pages} {58 } (\bibinfo {year} {2000})}\BibitemShut {NoStop}%
\bibitem [{\citenamefont {Starovoitova}\ \emph {et~al.}(2014)\citenamefont
  {Starovoitova}, \citenamefont {Tchelidze},\ and\ \citenamefont
  {Wells}}]{Starovoitova2014}%
  \BibitemOpen
  \bibfield  {author} {\bibinfo {author} {\bibfnamefont {V.~N.}\ \bibnamefont
  {Starovoitova}}, \bibinfo {author} {\bibfnamefont {L.}~\bibnamefont
  {Tchelidze}}, \ and\ \bibinfo {author} {\bibfnamefont {D.~P.}\ \bibnamefont
  {Wells}},\ }\href@noop {} {\bibfield  {journal} {\bibinfo  {journal} {Appl.
  Radiat. Isot.}\ }\textbf {\bibinfo {volume} {85}},\ \bibinfo {pages} {39 }
  (\bibinfo {year} {2014})}\BibitemShut {NoStop}%
\bibitem [{\citenamefont {Mang\'era}\ \emph {et~al.}(2015)\citenamefont
  {Mang\'era}, \citenamefont {Ogbomo}, \citenamefont {Zriba}, \citenamefont
  {Fitzpatrick}, \citenamefont {Brown}, \citenamefont {Pellerin}, \citenamefont
  {Barnard}, \citenamefont {Saunders},\ and\ \citenamefont
  {de~Jong}}]{Mangera2015}%
  \BibitemOpen
  \bibfield  {author} {\bibinfo {author} {\bibfnamefont {K.}~\bibnamefont
  {Mang\'era}}, \bibinfo {author} {\bibfnamefont {K.}~\bibnamefont {Ogbomo}},
  \bibinfo {author} {\bibfnamefont {R.}~\bibnamefont {Zriba}}, \bibinfo
  {author} {\bibfnamefont {J.}~\bibnamefont {Fitzpatrick}}, \bibinfo {author}
  {\bibfnamefont {J.}~\bibnamefont {Brown}}, \bibinfo {author} {\bibfnamefont
  {E.}~\bibnamefont {Pellerin}}, \bibinfo {author} {\bibfnamefont
  {J.}~\bibnamefont {Barnard}}, \bibinfo {author} {\bibfnamefont
  {C.}~\bibnamefont {Saunders}}, \ and\ \bibinfo {author} {\bibfnamefont
  {M.}~\bibnamefont {de~Jong}},\ }\href@noop {} {\bibfield  {journal} {\bibinfo
   {journal} {J. Radioanal. Nucl. Chem.}\ }\textbf {\bibinfo {volume} {305}},\
  \bibinfo {pages} {79} (\bibinfo {year} {2015})}\BibitemShut {NoStop}%
\bibitem [{\citenamefont {Fujiwara}\ \emph {et~al.}(2017)\citenamefont
  {Fujiwara}, \citenamefont {Nakai}, \citenamefont {Takahashi}, \citenamefont
  {Hayakawa}, \citenamefont {Shizuma}, \citenamefont {Miyamoto}, \citenamefont
  {Fan}, \citenamefont {Takemoto}, \citenamefont {Yamaguchi},\ and\
  \citenamefont {Nishimura}}]{Fujiwara2017}%
  \BibitemOpen
  \bibfield  {author} {\bibinfo {author} {\bibfnamefont {M.}~\bibnamefont
  {Fujiwara}}, \bibinfo {author} {\bibfnamefont {K.}~\bibnamefont {Nakai}},
  \bibinfo {author} {\bibfnamefont {N.}~\bibnamefont {Takahashi}}, \bibinfo
  {author} {\bibfnamefont {T.}~\bibnamefont {Hayakawa}}, \bibinfo {author}
  {\bibfnamefont {T.}~\bibnamefont {Shizuma}}, \bibinfo {author} {\bibfnamefont
  {S.}~\bibnamefont {Miyamoto}}, \bibinfo {author} {\bibfnamefont {G.~T.}\
  \bibnamefont {Fan}}, \bibinfo {author} {\bibfnamefont {A.}~\bibnamefont
  {Takemoto}}, \bibinfo {author} {\bibfnamefont {M.}~\bibnamefont {Yamaguchi}},
  \ and\ \bibinfo {author} {\bibfnamefont {M.}~\bibnamefont {Nishimura}},\
  }\href@noop {} {\bibfield  {journal} {\bibinfo  {journal} {Phys. Part.
  Nucl.}\ }\textbf {\bibinfo {volume} {48}},\ \bibinfo {pages} {124} (\bibinfo
  {year} {2017})}\BibitemShut {NoStop}%
\bibitem [{\citenamefont {Dikiy}\ \emph {et~al.}(2015)\citenamefont {Dikiy},
  \citenamefont {Dovbnya}, \citenamefont {Lyashko}, \citenamefont {Medvedeva},
  \citenamefont {Medvedev}, \citenamefont {Uvarov}, \citenamefont {Fedorets},\
  and\ \citenamefont {Krasnoselskiy}}]{Dikiy-nano}%
  \BibitemOpen
  \bibfield  {author} {\bibinfo {author} {\bibfnamefont {N.}~\bibnamefont
  {Dikiy}}, \bibinfo {author} {\bibfnamefont {A.}~\bibnamefont {Dovbnya}},
  \bibinfo {author} {\bibfnamefont {Y.}~\bibnamefont {Lyashko}}, \bibinfo
  {author} {\bibfnamefont {E.}~\bibnamefont {Medvedeva}}, \bibinfo {author}
  {\bibfnamefont {D.}~\bibnamefont {Medvedev}}, \bibinfo {author}
  {\bibfnamefont {V.}~\bibnamefont {Uvarov}}, \bibinfo {author} {\bibfnamefont
  {I.}~\bibnamefont {Fedorets}}, \ and\ \bibinfo {author} {\bibfnamefont
  {N.}~\bibnamefont {Krasnoselskiy}},\ }in\ \href@noop {} {\emph {\bibinfo
  {booktitle} {Proc. XXIV Int. Conf. on Charged Particle Accelerators}}}\
  (\bibinfo {address} {Kharkiv, Ukraine},\ \bibinfo {year} {2015})\ p.~\bibinfo
  {pages} {70}\BibitemShut {NoStop}%
\bibitem [{\citenamefont {Shibata}\ \emph {et~al.}(2011)\citenamefont
  {Shibata}, \citenamefont {Iwamoto}, \citenamefont {Nakagawa}, \citenamefont
  {Iwamoto}, \citenamefont {Ichihara}, \citenamefont {Kunieda}, \citenamefont
  {Chiba}, \citenamefont {Furutaka}, \citenamefont {Otuka}, \citenamefont
  {Ohsawa}, \citenamefont {Murata}, \citenamefont {Matsunobu}, \citenamefont
  {Zukeran}, \citenamefont {so~Kamada},\ and\ \citenamefont {ichi
  Katakura}}]{SHIBATA2011}%
  \BibitemOpen
  \bibfield  {author} {\bibinfo {author} {\bibfnamefont {K.}~\bibnamefont
  {Shibata}}, \bibinfo {author} {\bibfnamefont {O.}~\bibnamefont {Iwamoto}},
  \bibinfo {author} {\bibfnamefont {T.}~\bibnamefont {Nakagawa}}, \bibinfo
  {author} {\bibfnamefont {N.}~\bibnamefont {Iwamoto}}, \bibinfo {author}
  {\bibfnamefont {A.}~\bibnamefont {Ichihara}}, \bibinfo {author}
  {\bibfnamefont {S.}~\bibnamefont {Kunieda}}, \bibinfo {author} {\bibfnamefont
  {S.}~\bibnamefont {Chiba}}, \bibinfo {author} {\bibfnamefont
  {K.}~\bibnamefont {Furutaka}}, \bibinfo {author} {\bibfnamefont
  {N.}~\bibnamefont {Otuka}}, \bibinfo {author} {\bibfnamefont
  {T.}~\bibnamefont {Ohsawa}}, \bibinfo {author} {\bibfnamefont
  {T.}~\bibnamefont {Murata}}, \bibinfo {author} {\bibfnamefont
  {H.}~\bibnamefont {Matsunobu}}, \bibinfo {author} {\bibfnamefont
  {A.}~\bibnamefont {Zukeran}}, \bibinfo {author} {\bibnamefont {so~Kamada}}, \
  and\ \bibinfo {author} {\bibfnamefont {J.}~\bibnamefont {ichi Katakura}},\
  }\href {\doibase 10.1080/18811248.2011.9711675} {\bibfield  {journal}
  {\bibinfo  {journal} {Journal of Nuclear Science and Technology}\ }\textbf
  {\bibinfo {volume} {48}},\ \bibinfo {pages} {1} (\bibinfo {year} {2011})},\
  \Eprint {http://arxiv.org/abs/https://doi.org/10.1080/18811248.2011.9711675}
  {https://doi.org/10.1080/18811248.2011.9711675} \BibitemShut {NoStop}%
\bibitem [{\citenamefont {Koning}\ \emph {et~al.}(2006)\citenamefont {Koning},
  \citenamefont {Forrest}, \citenamefont {Kellett}, \citenamefont {Mills},
  \citenamefont {Henriksson}, \citenamefont {Rugama}, \citenamefont
  {Bersillon}, \citenamefont {Bouland}, \citenamefont {Courcelle},
  \citenamefont {Duijvestijn}, \citenamefont {Dupont}, \citenamefont {Kopecky},
  \citenamefont {Leichtle}, \citenamefont {Marie}, \citenamefont {Mattes},
  \citenamefont {Menapace}, \citenamefont {Morillon}, \citenamefont {Mounier},
  \citenamefont {Noguerre}, \citenamefont {Pereslavtsev}, \citenamefont
  {Romain}, \citenamefont {Serot}, \citenamefont {Simakov}, \citenamefont
  {Tagesen}, \citenamefont {Vonach}, \citenamefont {Batistoni}, \citenamefont
  {Bem}, \citenamefont {Gunsing}, \citenamefont {Pillon}, \citenamefont
  {Plompen}, \citenamefont {Rullhusen}, \citenamefont {Seidel}, \citenamefont
  {Avrigeanu}, \citenamefont {Avrigeanu}, \citenamefont {Bauge}, \citenamefont
  {Leeb}, \citenamefont {Lopez~Jimenez}, \citenamefont {Bernard}, \citenamefont
  {Bidaud}, \citenamefont {Dagan}, \citenamefont {Dean}, \citenamefont
  {Dos-Santos-Uzarralde}, \citenamefont {Fischer}, \citenamefont {Hogenbirk},
  \citenamefont {Jacqmin}, \citenamefont {Jouanne}, \citenamefont {Kodeli},
  \citenamefont {Leppanen}, \citenamefont {Marck}, \citenamefont {Perel},
  \citenamefont {Perry}, \citenamefont {Pescarini}, \citenamefont
  {Santamarina}, \citenamefont {Sublet}, \citenamefont {Trkov}, \citenamefont
  {Be}, \citenamefont {Huynh}, \citenamefont {Kellett}, \citenamefont {Mills},
  \citenamefont {Nichols}, \citenamefont {Henriksson}, \citenamefont
  {Nordborg}, \citenamefont {Nouri}, \citenamefont {Rugama},\ and\
  \citenamefont {Sartori}}]{Koning2006}%
  \BibitemOpen
  \bibfield  {author} {\bibinfo {author} {\bibfnamefont {A.}~\bibnamefont
  {Koning}}, \bibinfo {author} {\bibfnamefont {R.}~\bibnamefont {Forrest}},
  \bibinfo {author} {\bibfnamefont {M.}~\bibnamefont {Kellett}}, \bibinfo
  {author} {\bibfnamefont {R.}~\bibnamefont {Mills}}, \bibinfo {author}
  {\bibfnamefont {H.}~\bibnamefont {Henriksson}}, \bibinfo {author}
  {\bibfnamefont {Y.}~\bibnamefont {Rugama}}, \bibinfo {author} {\bibfnamefont
  {O.}~\bibnamefont {Bersillon}}, \bibinfo {author} {\bibfnamefont
  {O.}~\bibnamefont {Bouland}}, \bibinfo {author} {\bibfnamefont
  {A.}~\bibnamefont {Courcelle}}, \bibinfo {author} {\bibfnamefont {M.~C.}\
  \bibnamefont {Duijvestijn}}, \bibinfo {author} {\bibfnamefont
  {E.}~\bibnamefont {Dupont}}, \bibinfo {author} {\bibfnamefont
  {J.}~\bibnamefont {Kopecky}}, \bibinfo {author} {\bibfnamefont
  {D.}~\bibnamefont {Leichtle}}, \bibinfo {author} {\bibfnamefont
  {F.}~\bibnamefont {Marie}}, \bibinfo {author} {\bibfnamefont
  {M.}~\bibnamefont {Mattes}}, \bibinfo {author} {\bibfnamefont
  {E.}~\bibnamefont {Menapace}}, \bibinfo {author} {\bibfnamefont
  {B.}~\bibnamefont {Morillon}}, \bibinfo {author} {\bibfnamefont
  {C.}~\bibnamefont {Mounier}}, \bibinfo {author} {\bibfnamefont
  {G.}~\bibnamefont {Noguerre}}, \bibinfo {author} {\bibfnamefont
  {P.}~\bibnamefont {Pereslavtsev}}, \bibinfo {author} {\bibfnamefont
  {P.}~\bibnamefont {Romain}}, \bibinfo {author} {\bibfnamefont
  {O.}~\bibnamefont {Serot}}, \bibinfo {author} {\bibfnamefont
  {S.}~\bibnamefont {Simakov}}, \bibinfo {author} {\bibfnamefont
  {S.}~\bibnamefont {Tagesen}}, \bibinfo {author} {\bibfnamefont
  {H.}~\bibnamefont {Vonach}}, \bibinfo {author} {\bibfnamefont
  {P.}~\bibnamefont {Batistoni}}, \bibinfo {author} {\bibfnamefont
  {P.}~\bibnamefont {Bem}}, \bibinfo {author} {\bibfnamefont {F.}~\bibnamefont
  {Gunsing}}, \bibinfo {author} {\bibfnamefont {M.}~\bibnamefont {Pillon}},
  \bibinfo {author} {\bibfnamefont {A.}~\bibnamefont {Plompen}}, \bibinfo
  {author} {\bibfnamefont {P.}~\bibnamefont {Rullhusen}}, \bibinfo {author}
  {\bibfnamefont {K.}~\bibnamefont {Seidel}}, \bibinfo {author} {\bibfnamefont
  {M.}~\bibnamefont {Avrigeanu}}, \bibinfo {author} {\bibfnamefont
  {V.}~\bibnamefont {Avrigeanu}}, \bibinfo {author} {\bibfnamefont
  {E.}~\bibnamefont {Bauge}}, \bibinfo {author} {\bibfnamefont
  {H.}~\bibnamefont {Leeb}}, \bibinfo {author} {\bibfnamefont {M.~J.}\
  \bibnamefont {Lopez~Jimenez}}, \bibinfo {author} {\bibfnamefont
  {D.}~\bibnamefont {Bernard}}, \bibinfo {author} {\bibfnamefont
  {A.}~\bibnamefont {Bidaud}}, \bibinfo {author} {\bibfnamefont
  {R.}~\bibnamefont {Dagan}}, \bibinfo {author} {\bibfnamefont
  {C.}~\bibnamefont {Dean}}, \bibinfo {author} {\bibfnamefont {P.}~\bibnamefont
  {Dos-Santos-Uzarralde}}, \bibinfo {author} {\bibfnamefont {U.}~\bibnamefont
  {Fischer}}, \bibinfo {author} {\bibfnamefont {A.}~\bibnamefont {Hogenbirk}},
  \bibinfo {author} {\bibfnamefont {R.}~\bibnamefont {Jacqmin}}, \bibinfo
  {author} {\bibfnamefont {C.}~\bibnamefont {Jouanne}}, \bibinfo {author}
  {\bibfnamefont {I.}~\bibnamefont {Kodeli}}, \bibinfo {author} {\bibfnamefont
  {J.}~\bibnamefont {Leppanen}}, \bibinfo {author} {\bibfnamefont {S.~C.
  v.~d.}\ \bibnamefont {Marck}}, \bibinfo {author} {\bibfnamefont
  {R.}~\bibnamefont {Perel}}, \bibinfo {author} {\bibfnamefont
  {R.}~\bibnamefont {Perry}}, \bibinfo {author} {\bibfnamefont
  {M.}~\bibnamefont {Pescarini}}, \bibinfo {author} {\bibfnamefont
  {A.}~\bibnamefont {Santamarina}}, \bibinfo {author} {\bibfnamefont {J.~C.}\
  \bibnamefont {Sublet}}, \bibinfo {author} {\bibfnamefont {A.}~\bibnamefont
  {Trkov}}, \bibinfo {author} {\bibfnamefont {M.~M.}\ \bibnamefont {Be}},
  \bibinfo {author} {\bibfnamefont {T.~D.}\ \bibnamefont {Huynh}}, \bibinfo
  {author} {\bibfnamefont {M.~A.}\ \bibnamefont {Kellett}}, \bibinfo {author}
  {\bibfnamefont {R.}~\bibnamefont {Mills}}, \bibinfo {author} {\bibfnamefont
  {A.}~\bibnamefont {Nichols}}, \bibinfo {author} {\bibfnamefont
  {H.}~\bibnamefont {Henriksson}}, \bibinfo {author} {\bibfnamefont
  {C.}~\bibnamefont {Nordborg}}, \bibinfo {author} {\bibfnamefont
  {A.}~\bibnamefont {Nouri}}, \bibinfo {author} {\bibfnamefont
  {Y.}~\bibnamefont {Rugama}}, \ and\ \bibinfo {author} {\bibfnamefont
  {E.}~\bibnamefont {Sartori}},\ }\href@noop {} {\emph {\bibinfo {title} {The
  JEFF-31 Nuclear Data Library - JEFF Report 21}}},\ \bibinfo {type} {Tech.
  Rep.}\ (\bibinfo {address} {Nuclear Energy Agency of the OECD (NEA)},\
  \bibinfo {year} {2006})\BibitemShut {NoStop}%
\bibitem [{\citenamefont {Chadwick}\ \emph {et~al.}(2011)\citenamefont
  {Chadwick}, \citenamefont {Herman}, \citenamefont {Oblozinsky}, \citenamefont
  {Dunn}, \citenamefont {Danon}, \citenamefont {Kahler}, \citenamefont {Smith},
  \citenamefont {Pritychenko}, \citenamefont {Arbanas}, \citenamefont
  {Arcilla}, \citenamefont {Brewer}, \citenamefont {Brown}, \citenamefont
  {Capote}, \citenamefont {Carlson}, \citenamefont {Cho}, \citenamefont
  {Derrien}, \citenamefont {Guber}, \citenamefont {Hale}, \citenamefont
  {Hoblit}, \citenamefont {Holloway}, \citenamefont {Johnson}, \citenamefont
  {Kawano}, \citenamefont {Kiedrowski}, \citenamefont {Kim}, \citenamefont
  {Kunieda}, \citenamefont {Larson}, \citenamefont {Leal}, \citenamefont
  {Lestone}, \citenamefont {Little}, \citenamefont {McCutchan}, \citenamefont
  {MacFarlane}, \citenamefont {MacInnes}, \citenamefont {Mattoon},
  \citenamefont {McKnight}, \citenamefont {Mughabghab}, \citenamefont {Nobre},
  \citenamefont {Palmiotti}, \citenamefont {Palumbo}, \citenamefont {Pigni},
  \citenamefont {Pronyaev}, \citenamefont {Sayer}, \citenamefont {Sonzogni},
  \citenamefont {Summers}, \citenamefont {Talou}, \citenamefont {Thompson},
  \citenamefont {Trkov}, \citenamefont {Vogt}, \citenamefont {van~der Marck},
  \citenamefont {Wallner}, \citenamefont {White}, \citenamefont {Wiarda},\ and\
  \citenamefont {Young}}]{Chadwick2011}%
  \BibitemOpen
  \bibfield  {author} {\bibinfo {author} {\bibfnamefont {M.}~\bibnamefont
  {Chadwick}}, \bibinfo {author} {\bibfnamefont {M.}~\bibnamefont {Herman}},
  \bibinfo {author} {\bibfnamefont {P.}~\bibnamefont {Oblozinsky}}, \bibinfo
  {author} {\bibfnamefont {M.}~\bibnamefont {Dunn}}, \bibinfo {author}
  {\bibfnamefont {Y.}~\bibnamefont {Danon}}, \bibinfo {author} {\bibfnamefont
  {A.}~\bibnamefont {Kahler}}, \bibinfo {author} {\bibfnamefont
  {D.}~\bibnamefont {Smith}}, \bibinfo {author} {\bibfnamefont
  {B.}~\bibnamefont {Pritychenko}}, \bibinfo {author} {\bibfnamefont
  {G.}~\bibnamefont {Arbanas}}, \bibinfo {author} {\bibfnamefont
  {R.}~\bibnamefont {Arcilla}}, \bibinfo {author} {\bibfnamefont
  {R.}~\bibnamefont {Brewer}}, \bibinfo {author} {\bibfnamefont
  {D.}~\bibnamefont {Brown}}, \bibinfo {author} {\bibfnamefont
  {R.}~\bibnamefont {Capote}}, \bibinfo {author} {\bibfnamefont
  {A.}~\bibnamefont {Carlson}}, \bibinfo {author} {\bibfnamefont
  {Y.}~\bibnamefont {Cho}}, \bibinfo {author} {\bibfnamefont {H.}~\bibnamefont
  {Derrien}}, \bibinfo {author} {\bibfnamefont {K.}~\bibnamefont {Guber}},
  \bibinfo {author} {\bibfnamefont {G.}~\bibnamefont {Hale}}, \bibinfo {author}
  {\bibfnamefont {S.}~\bibnamefont {Hoblit}}, \bibinfo {author} {\bibfnamefont
  {S.}~\bibnamefont {Holloway}}, \bibinfo {author} {\bibfnamefont
  {T.}~\bibnamefont {Johnson}}, \bibinfo {author} {\bibfnamefont
  {T.}~\bibnamefont {Kawano}}, \bibinfo {author} {\bibfnamefont
  {B.}~\bibnamefont {Kiedrowski}}, \bibinfo {author} {\bibfnamefont
  {H.}~\bibnamefont {Kim}}, \bibinfo {author} {\bibfnamefont {S.}~\bibnamefont
  {Kunieda}}, \bibinfo {author} {\bibfnamefont {N.}~\bibnamefont {Larson}},
  \bibinfo {author} {\bibfnamefont {L.}~\bibnamefont {Leal}}, \bibinfo {author}
  {\bibfnamefont {J.}~\bibnamefont {Lestone}}, \bibinfo {author} {\bibfnamefont
  {R.}~\bibnamefont {Little}}, \bibinfo {author} {\bibfnamefont
  {E.}~\bibnamefont {McCutchan}}, \bibinfo {author} {\bibfnamefont
  {R.}~\bibnamefont {MacFarlane}}, \bibinfo {author} {\bibfnamefont
  {M.}~\bibnamefont {MacInnes}}, \bibinfo {author} {\bibfnamefont
  {C.}~\bibnamefont {Mattoon}}, \bibinfo {author} {\bibfnamefont
  {R.}~\bibnamefont {McKnight}}, \bibinfo {author} {\bibfnamefont
  {S.}~\bibnamefont {Mughabghab}}, \bibinfo {author} {\bibfnamefont
  {G.}~\bibnamefont {Nobre}}, \bibinfo {author} {\bibfnamefont
  {G.}~\bibnamefont {Palmiotti}}, \bibinfo {author} {\bibfnamefont
  {A.}~\bibnamefont {Palumbo}}, \bibinfo {author} {\bibfnamefont
  {M.}~\bibnamefont {Pigni}}, \bibinfo {author} {\bibfnamefont
  {V.}~\bibnamefont {Pronyaev}}, \bibinfo {author} {\bibfnamefont
  {R.}~\bibnamefont {Sayer}}, \bibinfo {author} {\bibfnamefont
  {A.}~\bibnamefont {Sonzogni}}, \bibinfo {author} {\bibfnamefont
  {N.}~\bibnamefont {Summers}}, \bibinfo {author} {\bibfnamefont
  {P.}~\bibnamefont {Talou}}, \bibinfo {author} {\bibfnamefont
  {I.}~\bibnamefont {Thompson}}, \bibinfo {author} {\bibfnamefont
  {A.}~\bibnamefont {Trkov}}, \bibinfo {author} {\bibfnamefont
  {R.}~\bibnamefont {Vogt}}, \bibinfo {author} {\bibfnamefont {S.}~\bibnamefont
  {van~der Marck}}, \bibinfo {author} {\bibfnamefont {A.}~\bibnamefont
  {Wallner}}, \bibinfo {author} {\bibfnamefont {M.}~\bibnamefont {White}},
  \bibinfo {author} {\bibfnamefont {D.}~\bibnamefont {Wiarda}}, \ and\ \bibinfo
  {author} {\bibfnamefont {P.}~\bibnamefont {Young}},\ }\href {\doibase
  https://doi.org/10.1016/j.nds.2011.11.002} {\bibfield  {journal} {\bibinfo
  {journal} {Nuclear Data Sheets}\ }\textbf {\bibinfo {volume} {112}},\
  \bibinfo {pages} {2887 } (\bibinfo {year} {2011})},\ \bibinfo {note} {special
  Issue on ENDF/B-VII.1 Library}\BibitemShut {NoStop}%
\bibitem [{\citenamefont {Beil}\ \emph {et~al.}(1974)\citenamefont {Beil},
  \citenamefont {Berg\'ere}, \citenamefont {Carlos}, \citenamefont
  {Lepr\^etre}, \citenamefont {Miniac},\ and\ \citenamefont
  {Veyssi\'ere}}]{Beil1974}%
  \BibitemOpen
  \bibfield  {author} {\bibinfo {author} {\bibfnamefont {H.}~\bibnamefont
  {Beil}}, \bibinfo {author} {\bibfnamefont {R.}~\bibnamefont {Berg\'ere}},
  \bibinfo {author} {\bibfnamefont {P.}~\bibnamefont {Carlos}}, \bibinfo
  {author} {\bibfnamefont {A.}~\bibnamefont {Lepr\^etre}}, \bibinfo {author}
  {\bibfnamefont {A.~D.}\ \bibnamefont {Miniac}}, \ and\ \bibinfo {author}
  {\bibfnamefont {A.}~\bibnamefont {Veyssi\'ere}},\ }\href@noop {} {\bibfield
  {journal} {\bibinfo  {journal} {Nucl. Phys. A}\ }\textbf {\bibinfo {volume}
  {227}},\ \bibinfo {pages} {427 } (\bibinfo {year} {1974})}\BibitemShut
  {NoStop}%
\bibitem [{\citenamefont {Červenák}\ and\ \citenamefont
  {Lebeda}(2016)}]{CERVENAK201632}%
  \BibitemOpen
  \bibfield  {author} {\bibinfo {author} {\bibfnamefont {J.}~\bibnamefont
  {Červenák}}\ and\ \bibinfo {author} {\bibfnamefont {O.}~\bibnamefont
  {Lebeda}},\ }\href {\doibase https://doi.org/10.1016/j.nimb.2016.05.006}
  {\bibfield  {journal} {\bibinfo  {journal} {Nuclear Instruments and Methods
  in Physics Research Section B: Beam Interactions with Materials and Atoms}\
  }\textbf {\bibinfo {volume} {380}},\ \bibinfo {pages} {32 } (\bibinfo {year}
  {2016})}\BibitemShut {NoStop}%
\bibitem [{\citenamefont {Dikiy}\ \emph {et~al.}(2016)\citenamefont {Dikiy},
  \citenamefont {Dovbnya}, \citenamefont {Fedorchenko},\ and\ \citenamefont
  {Khazhmuradov}}]{Dikiy2016}%
  \BibitemOpen
  \bibfield  {author} {\bibinfo {author} {\bibfnamefont {N.}~\bibnamefont
  {Dikiy}}, \bibinfo {author} {\bibfnamefont {A.}~\bibnamefont {Dovbnya}},
  \bibinfo {author} {\bibfnamefont {D.}~\bibnamefont {Fedorchenko}}, \ and\
  \bibinfo {author} {\bibfnamefont {M.}~\bibnamefont {Khazhmuradov}},\
  }\href@noop {} {\bibfield  {journal} {\bibinfo  {journal} {Appl. Radiat.
  Isot.}\ }\textbf {\bibinfo {volume} {114}},\ \bibinfo {pages} {7 } (\bibinfo
  {year} {2016})}\BibitemShut {NoStop}%
\bibitem [{\citenamefont {Agostinelli}\ \emph {et~al.}(2003)\citenamefont
  {Agostinelli}, \citenamefont {Allison}, \citenamefont {Amako}, \citenamefont
  {Apostolakis}, \citenamefont {Araujo}, \citenamefont {Arce}, \citenamefont
  {Asai}, \citenamefont {Axen}, \citenamefont {Banerjee}, \citenamefont
  {Barrand}, \citenamefont {Behner}, \citenamefont {Bellagamba}, \citenamefont
  {Boudreau}, \citenamefont {Broglia}, \citenamefont {Brunengo}, \citenamefont
  {Burkhardt}, \citenamefont {Chauvie}, \citenamefont {Chuma}, \citenamefont
  {Chytracek}, \citenamefont {Cooperman}, \citenamefont {Cosmo}, \citenamefont
  {Degtyarenko}, \citenamefont {Dell'Acqua}, \citenamefont {Depaola},
  \citenamefont {Dietrich}, \citenamefont {Enami}, \citenamefont {Feliciello},
  \citenamefont {Ferguson}, \citenamefont {Fesefeldt}, \citenamefont {Folger},
  \citenamefont {Foppiano}, \citenamefont {Forti}, \citenamefont {Garelli},
  \citenamefont {Giani}, \citenamefont {Giannitrapani}, \citenamefont {Gibin},
  \citenamefont {Cadenas}, \citenamefont {Gonz\'alez}, \citenamefont {Abril},
  \citenamefont {Greeniaus}, \citenamefont {Greiner}, \citenamefont {Grichine},
  \citenamefont {Grossheim}, \citenamefont {Guatelli}, \citenamefont
  {Gumplinger}, \citenamefont {Hamatsu}, \citenamefont {Hashimoto},
  \citenamefont {Hasui}, \citenamefont {Heikkinen}, \citenamefont {Howard},
  \citenamefont {Ivanchenko}, \citenamefont {Johnson}, \citenamefont {Jones},
  \citenamefont {Kallenbach}, \citenamefont {Kanaya}, \citenamefont {Kawabata},
  \citenamefont {Kawabata}, \citenamefont {Kawaguti}, \citenamefont {Kelner},
  \citenamefont {Kent}, \citenamefont {Kimura}, \citenamefont {Kodama},
  \citenamefont {Kokoulin}, \citenamefont {Kossov}, \citenamefont {Kurashige},
  \citenamefont {Lamanna}, \citenamefont {Lamp\'en}, \citenamefont {Lara},
  \citenamefont {Lefebure}, \citenamefont {Lei}, \citenamefont {Liendl},
  \citenamefont {Lockman}, \citenamefont {Longo}, \citenamefont {Magni},
  \citenamefont {Maire}, \citenamefont {Medernach}, \citenamefont {Minamimoto},
  \citenamefont {de~Freitas}, \citenamefont {Morita}, \citenamefont {Murakami},
  \citenamefont {Nagamatu}, \citenamefont {Nartallo}, \citenamefont {Nieminen},
  \citenamefont {Nishimura}, \citenamefont {Ohtsubo}, \citenamefont {Okamura},
  \citenamefont {O'Neale}, \citenamefont {Oohata}, \citenamefont {Paech},
  \citenamefont {Perl}, \citenamefont {Pfeiffer}, \citenamefont {Pia},
  \citenamefont {Ranjard}, \citenamefont {Rybin}, \citenamefont {Sadilov},
  \citenamefont {Salvo}, \citenamefont {Santin}, \citenamefont {Sasaki},
  \citenamefont {Savvas}, \citenamefont {Sawada}, \citenamefont {Scherer},
  \citenamefont {Sei}, \citenamefont {Sirotenko}, \citenamefont {Smith},
  \citenamefont {Starkov}, \citenamefont {Stoecker}, \citenamefont {Sulkimo},
  \citenamefont {Takahata}, \citenamefont {Tanaka}, \citenamefont {Tcherniaev},
  \citenamefont {Tehrani}, \citenamefont {Tropeano}, \citenamefont {Truscott},
  \citenamefont {Uno}, \citenamefont {Urban}, \citenamefont {Urban},
  \citenamefont {Verderi}, \citenamefont {Walkden}, \citenamefont {Wander},
  \citenamefont {Weber}, \citenamefont {Wellisch}, \citenamefont {Wenaus},
  \citenamefont {Williams}, \citenamefont {Wright}, \citenamefont {Yamada},
  \citenamefont {Yoshida},\ and\ \citenamefont {Zschiesche}}]{Geant1}%
  \BibitemOpen
  \bibfield  {author} {\bibinfo {author} {\bibfnamefont {S.}~\bibnamefont
  {Agostinelli}}, \bibinfo {author} {\bibfnamefont {J.}~\bibnamefont
  {Allison}}, \bibinfo {author} {\bibfnamefont {K.}~\bibnamefont {Amako}},
  \bibinfo {author} {\bibfnamefont {J.}~\bibnamefont {Apostolakis}}, \bibinfo
  {author} {\bibfnamefont {H.}~\bibnamefont {Araujo}}, \bibinfo {author}
  {\bibfnamefont {P.}~\bibnamefont {Arce}}, \bibinfo {author} {\bibfnamefont
  {M.}~\bibnamefont {Asai}}, \bibinfo {author} {\bibfnamefont {D.}~\bibnamefont
  {Axen}}, \bibinfo {author} {\bibfnamefont {S.}~\bibnamefont {Banerjee}},
  \bibinfo {author} {\bibfnamefont {G.}~\bibnamefont {Barrand}}, \bibinfo
  {author} {\bibfnamefont {F.}~\bibnamefont {Behner}}, \bibinfo {author}
  {\bibfnamefont {L.}~\bibnamefont {Bellagamba}}, \bibinfo {author}
  {\bibfnamefont {J.}~\bibnamefont {Boudreau}}, \bibinfo {author}
  {\bibfnamefont {L.}~\bibnamefont {Broglia}}, \bibinfo {author} {\bibfnamefont
  {A.}~\bibnamefont {Brunengo}}, \bibinfo {author} {\bibfnamefont
  {H.}~\bibnamefont {Burkhardt}}, \bibinfo {author} {\bibfnamefont
  {S.}~\bibnamefont {Chauvie}}, \bibinfo {author} {\bibfnamefont
  {J.}~\bibnamefont {Chuma}}, \bibinfo {author} {\bibfnamefont
  {R.}~\bibnamefont {Chytracek}}, \bibinfo {author} {\bibfnamefont
  {G.}~\bibnamefont {Cooperman}}, \bibinfo {author} {\bibfnamefont
  {G.}~\bibnamefont {Cosmo}}, \bibinfo {author} {\bibfnamefont
  {P.}~\bibnamefont {Degtyarenko}}, \bibinfo {author} {\bibfnamefont
  {A.}~\bibnamefont {Dell'Acqua}}, \bibinfo {author} {\bibfnamefont
  {G.}~\bibnamefont {Depaola}}, \bibinfo {author} {\bibfnamefont
  {D.}~\bibnamefont {Dietrich}}, \bibinfo {author} {\bibfnamefont
  {R.}~\bibnamefont {Enami}}, \bibinfo {author} {\bibfnamefont
  {A.}~\bibnamefont {Feliciello}}, \bibinfo {author} {\bibfnamefont
  {C.}~\bibnamefont {Ferguson}}, \bibinfo {author} {\bibfnamefont
  {H.}~\bibnamefont {Fesefeldt}}, \bibinfo {author} {\bibfnamefont
  {G.}~\bibnamefont {Folger}}, \bibinfo {author} {\bibfnamefont
  {F.}~\bibnamefont {Foppiano}}, \bibinfo {author} {\bibfnamefont
  {A.}~\bibnamefont {Forti}}, \bibinfo {author} {\bibfnamefont
  {S.}~\bibnamefont {Garelli}}, \bibinfo {author} {\bibfnamefont
  {S.}~\bibnamefont {Giani}}, \bibinfo {author} {\bibfnamefont
  {R.}~\bibnamefont {Giannitrapani}}, \bibinfo {author} {\bibfnamefont
  {D.}~\bibnamefont {Gibin}}, \bibinfo {author} {\bibfnamefont {J.~G.}\
  \bibnamefont {Cadenas}}, \bibinfo {author} {\bibfnamefont {I.}~\bibnamefont
  {Gonz\'alez}}, \bibinfo {author} {\bibfnamefont {G.~G.}\ \bibnamefont
  {Abril}}, \bibinfo {author} {\bibfnamefont {G.}~\bibnamefont {Greeniaus}},
  \bibinfo {author} {\bibfnamefont {W.}~\bibnamefont {Greiner}}, \bibinfo
  {author} {\bibfnamefont {V.}~\bibnamefont {Grichine}}, \bibinfo {author}
  {\bibfnamefont {A.}~\bibnamefont {Grossheim}}, \bibinfo {author}
  {\bibfnamefont {S.}~\bibnamefont {Guatelli}}, \bibinfo {author}
  {\bibfnamefont {P.}~\bibnamefont {Gumplinger}}, \bibinfo {author}
  {\bibfnamefont {R.}~\bibnamefont {Hamatsu}}, \bibinfo {author} {\bibfnamefont
  {K.}~\bibnamefont {Hashimoto}}, \bibinfo {author} {\bibfnamefont
  {H.}~\bibnamefont {Hasui}}, \bibinfo {author} {\bibfnamefont
  {A.}~\bibnamefont {Heikkinen}}, \bibinfo {author} {\bibfnamefont
  {A.}~\bibnamefont {Howard}}, \bibinfo {author} {\bibfnamefont
  {V.}~\bibnamefont {Ivanchenko}}, \bibinfo {author} {\bibfnamefont
  {A.}~\bibnamefont {Johnson}}, \bibinfo {author} {\bibfnamefont
  {F.}~\bibnamefont {Jones}}, \bibinfo {author} {\bibfnamefont
  {J.}~\bibnamefont {Kallenbach}}, \bibinfo {author} {\bibfnamefont
  {N.}~\bibnamefont {Kanaya}}, \bibinfo {author} {\bibfnamefont
  {M.}~\bibnamefont {Kawabata}}, \bibinfo {author} {\bibfnamefont
  {Y.}~\bibnamefont {Kawabata}}, \bibinfo {author} {\bibfnamefont
  {M.}~\bibnamefont {Kawaguti}}, \bibinfo {author} {\bibfnamefont
  {S.}~\bibnamefont {Kelner}}, \bibinfo {author} {\bibfnamefont
  {P.}~\bibnamefont {Kent}}, \bibinfo {author} {\bibfnamefont {A.}~\bibnamefont
  {Kimura}}, \bibinfo {author} {\bibfnamefont {T.}~\bibnamefont {Kodama}},
  \bibinfo {author} {\bibfnamefont {R.}~\bibnamefont {Kokoulin}}, \bibinfo
  {author} {\bibfnamefont {M.}~\bibnamefont {Kossov}}, \bibinfo {author}
  {\bibfnamefont {H.}~\bibnamefont {Kurashige}}, \bibinfo {author}
  {\bibfnamefont {E.}~\bibnamefont {Lamanna}}, \bibinfo {author} {\bibfnamefont
  {T.}~\bibnamefont {Lamp\'en}}, \bibinfo {author} {\bibfnamefont
  {V.}~\bibnamefont {Lara}}, \bibinfo {author} {\bibfnamefont {V.}~\bibnamefont
  {Lefebure}}, \bibinfo {author} {\bibfnamefont {F.}~\bibnamefont {Lei}},
  \bibinfo {author} {\bibfnamefont {M.}~\bibnamefont {Liendl}}, \bibinfo
  {author} {\bibfnamefont {W.}~\bibnamefont {Lockman}}, \bibinfo {author}
  {\bibfnamefont {F.}~\bibnamefont {Longo}}, \bibinfo {author} {\bibfnamefont
  {S.}~\bibnamefont {Magni}}, \bibinfo {author} {\bibfnamefont
  {M.}~\bibnamefont {Maire}}, \bibinfo {author} {\bibfnamefont
  {E.}~\bibnamefont {Medernach}}, \bibinfo {author} {\bibfnamefont
  {K.}~\bibnamefont {Minamimoto}}, \bibinfo {author} {\bibfnamefont {P.~M.}\
  \bibnamefont {de~Freitas}}, \bibinfo {author} {\bibfnamefont
  {Y.}~\bibnamefont {Morita}}, \bibinfo {author} {\bibfnamefont
  {K.}~\bibnamefont {Murakami}}, \bibinfo {author} {\bibfnamefont
  {M.}~\bibnamefont {Nagamatu}}, \bibinfo {author} {\bibfnamefont
  {R.}~\bibnamefont {Nartallo}}, \bibinfo {author} {\bibfnamefont
  {P.}~\bibnamefont {Nieminen}}, \bibinfo {author} {\bibfnamefont
  {T.}~\bibnamefont {Nishimura}}, \bibinfo {author} {\bibfnamefont
  {K.}~\bibnamefont {Ohtsubo}}, \bibinfo {author} {\bibfnamefont
  {M.}~\bibnamefont {Okamura}}, \bibinfo {author} {\bibfnamefont
  {S.}~\bibnamefont {O'Neale}}, \bibinfo {author} {\bibfnamefont
  {Y.}~\bibnamefont {Oohata}}, \bibinfo {author} {\bibfnamefont
  {K.}~\bibnamefont {Paech}}, \bibinfo {author} {\bibfnamefont
  {J.}~\bibnamefont {Perl}}, \bibinfo {author} {\bibfnamefont {A.}~\bibnamefont
  {Pfeiffer}}, \bibinfo {author} {\bibfnamefont {M.}~\bibnamefont {Pia}},
  \bibinfo {author} {\bibfnamefont {F.}~\bibnamefont {Ranjard}}, \bibinfo
  {author} {\bibfnamefont {A.}~\bibnamefont {Rybin}}, \bibinfo {author}
  {\bibfnamefont {S.}~\bibnamefont {Sadilov}}, \bibinfo {author} {\bibfnamefont
  {E.~D.}\ \bibnamefont {Salvo}}, \bibinfo {author} {\bibfnamefont
  {G.}~\bibnamefont {Santin}}, \bibinfo {author} {\bibfnamefont
  {T.}~\bibnamefont {Sasaki}}, \bibinfo {author} {\bibfnamefont
  {N.}~\bibnamefont {Savvas}}, \bibinfo {author} {\bibfnamefont
  {Y.}~\bibnamefont {Sawada}}, \bibinfo {author} {\bibfnamefont
  {S.}~\bibnamefont {Scherer}}, \bibinfo {author} {\bibfnamefont
  {S.}~\bibnamefont {Sei}}, \bibinfo {author} {\bibfnamefont {V.}~\bibnamefont
  {Sirotenko}}, \bibinfo {author} {\bibfnamefont {D.}~\bibnamefont {Smith}},
  \bibinfo {author} {\bibfnamefont {N.}~\bibnamefont {Starkov}}, \bibinfo
  {author} {\bibfnamefont {H.}~\bibnamefont {Stoecker}}, \bibinfo {author}
  {\bibfnamefont {J.}~\bibnamefont {Sulkimo}}, \bibinfo {author} {\bibfnamefont
  {M.}~\bibnamefont {Takahata}}, \bibinfo {author} {\bibfnamefont
  {S.}~\bibnamefont {Tanaka}}, \bibinfo {author} {\bibfnamefont
  {E.}~\bibnamefont {Tcherniaev}}, \bibinfo {author} {\bibfnamefont {E.~S.}\
  \bibnamefont {Tehrani}}, \bibinfo {author} {\bibfnamefont {M.}~\bibnamefont
  {Tropeano}}, \bibinfo {author} {\bibfnamefont {P.}~\bibnamefont {Truscott}},
  \bibinfo {author} {\bibfnamefont {H.}~\bibnamefont {Uno}}, \bibinfo {author}
  {\bibfnamefont {L.}~\bibnamefont {Urban}}, \bibinfo {author} {\bibfnamefont
  {P.}~\bibnamefont {Urban}}, \bibinfo {author} {\bibfnamefont
  {M.}~\bibnamefont {Verderi}}, \bibinfo {author} {\bibfnamefont
  {A.}~\bibnamefont {Walkden}}, \bibinfo {author} {\bibfnamefont
  {W.}~\bibnamefont {Wander}}, \bibinfo {author} {\bibfnamefont
  {H.}~\bibnamefont {Weber}}, \bibinfo {author} {\bibfnamefont
  {J.}~\bibnamefont {Wellisch}}, \bibinfo {author} {\bibfnamefont
  {T.}~\bibnamefont {Wenaus}}, \bibinfo {author} {\bibfnamefont
  {D.}~\bibnamefont {Williams}}, \bibinfo {author} {\bibfnamefont
  {D.}~\bibnamefont {Wright}}, \bibinfo {author} {\bibfnamefont
  {T.}~\bibnamefont {Yamada}}, \bibinfo {author} {\bibfnamefont
  {H.}~\bibnamefont {Yoshida}}, \ and\ \bibinfo {author} {\bibfnamefont
  {D.}~\bibnamefont {Zschiesche}},\ }\href@noop {} {\bibfield  {journal}
  {\bibinfo  {journal} {Nucl. Instrum. Methods Phys. Res., Sect. A}\ }\textbf
  {\bibinfo {volume} {506}},\ \bibinfo {pages} {250 } (\bibinfo {year}
  {2003})}\BibitemShut {NoStop}%
\bibitem [{\citenamefont {Allison}\ \emph {et~al.}(2006)\citenamefont
  {Allison}, \citenamefont {Amako}, \citenamefont {Apostolakis}, \citenamefont
  {Araujo}, \citenamefont {Dubois}, \citenamefont {Asai}, \citenamefont
  {Barrand}, \citenamefont {Capra}, \citenamefont {Chauvie}, \citenamefont
  {Chytracek}, \citenamefont {Cirrone}, \citenamefont {Cooperman},
  \citenamefont {Cosmo}, \citenamefont {Cuttone}, \citenamefont {Daquino},
  \citenamefont {Donszelmann}, \citenamefont {Dressel}, \citenamefont {Folger},
  \citenamefont {Foppiano}, \citenamefont {Generowicz}, \citenamefont
  {Grichine}, \citenamefont {Guatelli}, \citenamefont {Gumplinger},
  \citenamefont {Heikkinen}, \citenamefont {Hrivnacova}, \citenamefont
  {Howard}, \citenamefont {Incerti}, \citenamefont {Ivanchenko}, \citenamefont
  {Johnson}, \citenamefont {Jones}, \citenamefont {Koi}, \citenamefont
  {Kokoulin}, \citenamefont {Kossov}, \citenamefont {Kurashige}, \citenamefont
  {Lara}, \citenamefont {Larsson}, \citenamefont {Lei}, \citenamefont {Link},
  \citenamefont {Longo}, \citenamefont {Maire}, \citenamefont {Mantero},
  \citenamefont {Mascialino}, \citenamefont {McLaren}, \citenamefont {Lorenzo},
  \citenamefont {Minamimoto}, \citenamefont {Murakami}, \citenamefont
  {Nieminen}, \citenamefont {Pandola}, \citenamefont {Parlati}, \citenamefont
  {Peralta}, \citenamefont {Perl}, \citenamefont {Pfeiffer}, \citenamefont
  {Pia}, \citenamefont {Ribon}, \citenamefont {Rodrigues}, \citenamefont
  {Russo}, \citenamefont {Sadilov}, \citenamefont {Santin}, \citenamefont
  {Sasaki}, \citenamefont {Smith}, \citenamefont {Starkov}, \citenamefont
  {Tanaka}, \citenamefont {Tcherniaev}, \citenamefont {Tome}, \citenamefont
  {Trindade}, \citenamefont {Truscott}, \citenamefont {Urban}, \citenamefont
  {Verderi}, \citenamefont {Walkden}, \citenamefont {Wellisch}, \citenamefont
  {Williams}, \citenamefont {Wright},\ and\ \citenamefont {Yoshida}}]{Geant2}%
  \BibitemOpen
  \bibfield  {author} {\bibinfo {author} {\bibfnamefont {J.}~\bibnamefont
  {Allison}}, \bibinfo {author} {\bibfnamefont {K.}~\bibnamefont {Amako}},
  \bibinfo {author} {\bibfnamefont {J.}~\bibnamefont {Apostolakis}}, \bibinfo
  {author} {\bibfnamefont {H.}~\bibnamefont {Araujo}}, \bibinfo {author}
  {\bibfnamefont {P.}~\bibnamefont {Dubois}}, \bibinfo {author} {\bibfnamefont
  {M.}~\bibnamefont {Asai}}, \bibinfo {author} {\bibfnamefont {G.}~\bibnamefont
  {Barrand}}, \bibinfo {author} {\bibfnamefont {R.}~\bibnamefont {Capra}},
  \bibinfo {author} {\bibfnamefont {S.}~\bibnamefont {Chauvie}}, \bibinfo
  {author} {\bibfnamefont {R.}~\bibnamefont {Chytracek}}, \bibinfo {author}
  {\bibfnamefont {G.}~\bibnamefont {Cirrone}}, \bibinfo {author} {\bibfnamefont
  {G.}~\bibnamefont {Cooperman}}, \bibinfo {author} {\bibfnamefont
  {G.}~\bibnamefont {Cosmo}}, \bibinfo {author} {\bibfnamefont
  {G.}~\bibnamefont {Cuttone}}, \bibinfo {author} {\bibfnamefont
  {G.}~\bibnamefont {Daquino}}, \bibinfo {author} {\bibfnamefont
  {M.}~\bibnamefont {Donszelmann}}, \bibinfo {author} {\bibfnamefont
  {M.}~\bibnamefont {Dressel}}, \bibinfo {author} {\bibfnamefont
  {G.}~\bibnamefont {Folger}}, \bibinfo {author} {\bibfnamefont
  {F.}~\bibnamefont {Foppiano}}, \bibinfo {author} {\bibfnamefont
  {J.}~\bibnamefont {Generowicz}}, \bibinfo {author} {\bibfnamefont
  {V.}~\bibnamefont {Grichine}}, \bibinfo {author} {\bibfnamefont
  {S.}~\bibnamefont {Guatelli}}, \bibinfo {author} {\bibfnamefont
  {P.}~\bibnamefont {Gumplinger}}, \bibinfo {author} {\bibfnamefont
  {A.}~\bibnamefont {Heikkinen}}, \bibinfo {author} {\bibfnamefont
  {I.}~\bibnamefont {Hrivnacova}}, \bibinfo {author} {\bibfnamefont
  {A.}~\bibnamefont {Howard}}, \bibinfo {author} {\bibfnamefont
  {S.}~\bibnamefont {Incerti}}, \bibinfo {author} {\bibfnamefont
  {V.}~\bibnamefont {Ivanchenko}}, \bibinfo {author} {\bibfnamefont
  {T.}~\bibnamefont {Johnson}}, \bibinfo {author} {\bibfnamefont
  {F.}~\bibnamefont {Jones}}, \bibinfo {author} {\bibfnamefont
  {T.}~\bibnamefont {Koi}}, \bibinfo {author} {\bibfnamefont {R.}~\bibnamefont
  {Kokoulin}}, \bibinfo {author} {\bibfnamefont {M.}~\bibnamefont {Kossov}},
  \bibinfo {author} {\bibfnamefont {H.}~\bibnamefont {Kurashige}}, \bibinfo
  {author} {\bibfnamefont {V.}~\bibnamefont {Lara}}, \bibinfo {author}
  {\bibfnamefont {S.}~\bibnamefont {Larsson}}, \bibinfo {author} {\bibfnamefont
  {F.}~\bibnamefont {Lei}}, \bibinfo {author} {\bibfnamefont {O.}~\bibnamefont
  {Link}}, \bibinfo {author} {\bibfnamefont {F.}~\bibnamefont {Longo}},
  \bibinfo {author} {\bibfnamefont {M.}~\bibnamefont {Maire}}, \bibinfo
  {author} {\bibfnamefont {A.}~\bibnamefont {Mantero}}, \bibinfo {author}
  {\bibfnamefont {B.}~\bibnamefont {Mascialino}}, \bibinfo {author}
  {\bibfnamefont {I.}~\bibnamefont {McLaren}}, \bibinfo {author} {\bibfnamefont
  {P.}~\bibnamefont {Lorenzo}}, \bibinfo {author} {\bibfnamefont
  {K.}~\bibnamefont {Minamimoto}}, \bibinfo {author} {\bibfnamefont
  {K.}~\bibnamefont {Murakami}}, \bibinfo {author} {\bibfnamefont
  {P.}~\bibnamefont {Nieminen}}, \bibinfo {author} {\bibfnamefont
  {L.}~\bibnamefont {Pandola}}, \bibinfo {author} {\bibfnamefont
  {S.}~\bibnamefont {Parlati}}, \bibinfo {author} {\bibfnamefont
  {L.}~\bibnamefont {Peralta}}, \bibinfo {author} {\bibfnamefont
  {J.}~\bibnamefont {Perl}}, \bibinfo {author} {\bibfnamefont {A.}~\bibnamefont
  {Pfeiffer}}, \bibinfo {author} {\bibfnamefont {M.}~\bibnamefont {Pia}},
  \bibinfo {author} {\bibfnamefont {A.}~\bibnamefont {Ribon}}, \bibinfo
  {author} {\bibfnamefont {P.}~\bibnamefont {Rodrigues}}, \bibinfo {author}
  {\bibfnamefont {G.}~\bibnamefont {Russo}}, \bibinfo {author} {\bibfnamefont
  {S.}~\bibnamefont {Sadilov}}, \bibinfo {author} {\bibfnamefont
  {G.}~\bibnamefont {Santin}}, \bibinfo {author} {\bibfnamefont
  {T.}~\bibnamefont {Sasaki}}, \bibinfo {author} {\bibfnamefont
  {D.}~\bibnamefont {Smith}}, \bibinfo {author} {\bibfnamefont
  {N.}~\bibnamefont {Starkov}}, \bibinfo {author} {\bibfnamefont
  {S.}~\bibnamefont {Tanaka}}, \bibinfo {author} {\bibfnamefont
  {E.}~\bibnamefont {Tcherniaev}}, \bibinfo {author} {\bibfnamefont
  {B.}~\bibnamefont {Tome}}, \bibinfo {author} {\bibfnamefont {A.}~\bibnamefont
  {Trindade}}, \bibinfo {author} {\bibfnamefont {P.}~\bibnamefont {Truscott}},
  \bibinfo {author} {\bibfnamefont {L.}~\bibnamefont {Urban}}, \bibinfo
  {author} {\bibfnamefont {M.}~\bibnamefont {Verderi}}, \bibinfo {author}
  {\bibfnamefont {A.}~\bibnamefont {Walkden}}, \bibinfo {author} {\bibfnamefont
  {J.}~\bibnamefont {Wellisch}}, \bibinfo {author} {\bibfnamefont
  {D.}~\bibnamefont {Williams}}, \bibinfo {author} {\bibfnamefont
  {D.}~\bibnamefont {Wright}}, \ and\ \bibinfo {author} {\bibfnamefont
  {H.}~\bibnamefont {Yoshida}},\ }\href {\doibase 10.1109/TNS.2006.869826}
  {\bibfield  {journal} {\bibinfo  {journal} {Nuclear Science, IEEE
  Transactions on}\ }\textbf {\bibinfo {volume} {53}},\ \bibinfo {pages} {270}
  (\bibinfo {year} {2006})}\BibitemShut {NoStop}%
\bibitem [{\citenamefont {Koning}\ \emph {et~al.}(2007)\citenamefont {Koning},
  \citenamefont {Hilaire},\ and\ \citenamefont {Duijvestijn}}]{Talys1}%
  \BibitemOpen
  \bibfield  {author} {\bibinfo {author} {\bibfnamefont {A.}~\bibnamefont
  {Koning}}, \bibinfo {author} {\bibfnamefont {S.}~\bibnamefont {Hilaire}}, \
  and\ \bibinfo {author} {\bibfnamefont {M.}~\bibnamefont {Duijvestijn}},\ }in\
  \href@noop {} {\emph {\bibinfo {booktitle} {Proc. Int. Conf. on Nuclear Data
  for Science and Technology}}}\ (\bibinfo {address} {Nice, France},\ \bibinfo
  {year} {2007})\ pp.\ \bibinfo {pages} {211--214}\BibitemShut {NoStop}%
\bibitem [{\citenamefont {Weisskopf}(1937)}]{Weisskopf1937}%
  \BibitemOpen
  \bibfield  {author} {\bibinfo {author} {\bibfnamefont {V.}~\bibnamefont
  {Weisskopf}},\ }\href@noop {} {\bibfield  {journal} {\bibinfo  {journal}
  {Phys. Rev.}\ }\textbf {\bibinfo {volume} {52}},\ \bibinfo {pages} {295}
  (\bibinfo {year} {1937})}\BibitemShut {NoStop}%
\bibitem [{\citenamefont {Dostrovsky}\ \emph {et~al.}(1959)\citenamefont
  {Dostrovsky}, \citenamefont {Fraenkel},\ and\ \citenamefont
  {Friedlander}}]{Dostrovsky1959}%
  \BibitemOpen
  \bibfield  {author} {\bibinfo {author} {\bibfnamefont {I.}~\bibnamefont
  {Dostrovsky}}, \bibinfo {author} {\bibfnamefont {Z.}~\bibnamefont
  {Fraenkel}}, \ and\ \bibinfo {author} {\bibfnamefont {G.}~\bibnamefont
  {Friedlander}},\ }\href@noop {} {\bibfield  {journal} {\bibinfo  {journal}
  {Phys. Rev.}\ }\textbf {\bibinfo {volume} {116}},\ \bibinfo {pages} {683}
  (\bibinfo {year} {1959})}\BibitemShut {NoStop}%
\bibitem [{\citenamefont {Mendenhall}\ and\ \citenamefont
  {Weller}(2005)}]{Mendenhall2005420}%
  \BibitemOpen
  \bibfield  {author} {\bibinfo {author} {\bibfnamefont {M.~H.}\ \bibnamefont
  {Mendenhall}}\ and\ \bibinfo {author} {\bibfnamefont {R.~A.}\ \bibnamefont
  {Weller}},\ }\href@noop {} {\bibfield  {journal} {\bibinfo  {journal} {Nucl.
  Instrum. Methods Phys. Res., Sect. B}\ }\textbf {\bibinfo {volume} {227}},\
  \bibinfo {pages} {420 } (\bibinfo {year} {2005})}\BibitemShut {NoStop}%
\end{thebibliography}%

\end{document}